\documentclass[lettersize,journal]{IEEEtran}
\usepackage{amsmath,amsfonts}
\usepackage{algorithmic}
\usepackage{algorithm}
\usepackage{array}
\usepackage[caption=false,font=normalsize,labelfont=sf,textfont=sf]{subfig}
\usepackage{textcomp}
\usepackage{stfloats}
\usepackage{url}
\usepackage{verbatim}
\usepackage{graphicx}
\usepackage{cite}
\begin{document}

\title{Genetic Algorithm-Based Inverse Design of Guided Wave Planar Terahertz Filters}

\author{
        Ali~Dehghanian$^{1,2}$,~\IEEEmembership{Student,~IEEE,}
        Thomas~Darcie$^{1}$,~\IEEEmembership{Fellow,~IEEE}
        and Levi~Smith$^{1,2,*}$,~\IEEEmembership{Member,~IEEE,}
\thanks{$^1$Department of Electrical and Computer Engineering, University of Victoria, Victoria, BC, V8P 5C2 Canada}
\thanks{$^2$Centre for Advanced Materials and Related Technology (CAMTEC), University of Victoria, 3800 Finnerty Rd, Victoria, BC, V8P 5C2, Canada.}
\thanks{$^*$Corresponding author: levismith@uvic.ca}
\thanks{This work was supported by a NSERC Discovery Grant.}
\thanks{Manuscript received May xx, xxxx; revised May xx, xxxx.}}

\markboth{IEEE Template (Preprint)}%
{Smith \MakeLowercase{\textit{et al.}}: IEEE Transactions on Terahertz Science and Technology}

\maketitle

\begin{abstract}
We present a genetic algorithm (GA)-based inverse design framework for synthesizing high-performance planar terahertz (THz) filters integrated with coplanar striplines (CPSs). The method efficiently explores high-dimensional design spaces to generate filter geometries matching user-defined S-parameter magnitude and phase responses, while enforcing structural connectivity for compatibility with terahertz system-on-chip (TSoC) platforms. To accelerate optimization, filter performance is evaluated using the ABCD matrix method, providing a significant computational advantage over full-wave simulations. Final validation is performed through finite element method (FEM) simulations. As a proof of concept, we design band-stop filters with center frequencies of 0.6, 0.8, and 1.0~THz, each with a 150~GHz target bandwidth, and demonstrate tunable rejection depths within a constant physical footprint. Optimization is guided by minimizing the root-mean-square error (RMSE) between simulated and target S-parameters. 
\end{abstract}


\section{Introduction}
\IEEEPARstart{T}{erahertz} (THz) filters are required for next-generation THz systems, allowing spectral control by selectively transmitting desired frequency bands while suppressing out-of-band signals. These functionalities are vital for a range of applications, including high-speed wireless communications, non-invasive medical imaging, and spectroscopic sensing~\cite{federici2010review, huang2011terahertz, jiang2024terahertz, gezimati2023terahertz, yan2022thz, fu2022applications}. The ability to engineer and tailor spectral responses with high accuracy directly impacts the efficiency, resolution, and functionality of THz devices across scientific, medical, and industrial domains.

Structurally, THz filters can be classified into planar and non-planar topologies, each offering specific advantages in terms of performance metrics such as insertion loss, bandwidth, and frequency selectivity, as well as fabrication compatibility and system-level integrability. Waveguide-based filters, such as rectangular metallic and parallel plate waveguides, are popular for their low insertion loss and effective mode confinement in conventional setups~\cite{gerhard2015comparative, lee2011terahertz}. Planar configurations, including coplanar waveguides (CPW) and coplanar striplines (CPS), are particularly attractive for on-chip applications due to their compatibility with standard photolithography and compact on-chip footprint~\cite{cabello2022capacitively,gomaa2020terahertz,dehghanian2025demonstration}.

Alternative filter designs, such as photonic crystal filters, exploit periodic dielectric structures to induce photonic bandgaps, offering precise spectral shaping capabilities~\cite{li2015tunable}. Metamaterial and metasurface filters, composed of engineered sub-wavelength resonant elements, provide enhanced design flexibility and enable tailored responses such as narrowband, broadband, and multi-band filtering~\cite{asl2020terahertz,horestani2013metamaterial}. Resonator-based filters, including ring and cavity resonators, are also widely used in integrated planar formats due to their high quality ($Q$) factors and small footprint~\cite{lin2019tunable,parker2021tunable}.


Traditionally, the design of THz filters has relied heavily on the designer’s expertise and theoretical models rooted in microwave and optical engineering. These conventional methods typically involve iterative tuning of structural parameters based on analytical frameworks such as transmission line theory and coupled-mode theory. While these models offer valuable physical insight, they are often limited in their applicability to geometrically complex or multi-objective structures. Consequently, the design process remains labor-intensive and constrained in its ability to efficiently explore the vast and high-dimensional design space, increasing the risk of overlooking high-performing or unconventional solutions. These limitations are particularly evident in the development of THz components such as filters, resonators, polarizers, and power splitters, where optimal performance often depends on intricate geometries and tightly coupled physical effects.

Here we adopt inverse design that was originally developed to optimize nanophotonic and electromagnetic structures beyond the capabilities of traditional analytical approaches. Early applications in photonic systems demonstrated its ability to generate compact, high-performance devices with non-intuitive geometries that satisfy specific optical transmission or field distribution requirements. In particular, its application to photonic integrated circuits (PICs) has facilitated the design of densely integrated components that offer enhanced functionality, compact footprints, and improved performance~\cite{molesky2018inverse}. The maturity and success of inverse design in silicon photonics not only validate its effectiveness but also highlight its potential for addressing analogous challenges in emerging THz integrated systems.


Several recent studies have explored inverse design strategies for optimizing THz components. Zhang et al.~\cite{zhang2022automatic} proposed a genetic algorithm (GA)-based inverse design framework to optimize dispersion profiles for broadband impedance matching. Their dual-metasurface absorber achieved 88\% absorption from 0.21 to 5~THz, demonstrating both high performance and computational efficiency. In a follow-up study, the same group used GA-assisted optimization to enhance THz metasurface design, using multiple objective functions to improve design reliability and automation~\cite{zhang2023accurate}.

Beyond GAs, machine learning-based approaches have been applied to accelerate and enhance the inverse design process. Li et al.~\cite{li2020applications} used artificial neural networks (ANNs) to accurately predict reflection spectra and optimize micro/nano THz metasurface structures with custom optical responses. Similarly, Mashayekhi et al.~\cite{mashayekhi2023reconfigurable} developed an ANN-assisted inverse-designed graphene-based absorber, achieving 96.33\% absorption across 0.5–10~THz and enabling rapid parameter selection for THz detection. Deep learning has also been extended to THz antenna design. Karahan et al.~\cite{karahan2024deep} introduced a deep neural network (DNN)-based inverse design approach for multi-band graphene patch antennas operating between 2–5~THz. Their model achieved 13 frequency bands, up to 8.8~dB gain, full 360° beam steering, and 93\% prediction accuracy, significantly accelerating the design cycle. Furthermore, Ding et al.~\cite{ding2024metasurface} proposed a Finite-Difference Time-Domain (FDTD)-based inverse design framework for a 3D-printable diffractive optical element (DOE) capable of THz spectral splitting between 0.5–0.7~THz. Their simulation and experimental results demonstrated a compact, low-cost solution for portable THz spectroscopy and communication.

Although inverse design methodologies have been successfully applied to various THz components, such as metasurfaces, absorbers, antennas, and filters, the efficient design of planar THz filters integrated with guided-wave transmission lines remains largely unexplored. Addressing this gap, we present a GA-based inverse design framework specifically tailored for planar guided-wave THz on-chip filters. Using evolutionary optimization strategies, the proposed method enables the systematic exploration of complex, high-dimensional design spaces and facilitates the realization of high-performance THz filters compatible with standard lithographic fabrication processes.

To ensure computational feasibility, filter performance is evaluated using the ABCD matrix method, which offers substantial acceleration compared to full-wave simulations while maintaining sufficient modeling accuracy. The final validation of the optimized designs is performed through finite element method (FEM) simulations. As a proof-of-concept, we demonstrate the inverse design of band-stop filters with varying center frequencies and rejection depths, all within a fixed device footprint. To the best of our knowledge, this is the first application of a GA-based inverse design approach that employs the ABCD matrix method for CPS-integrated THz filters, highlighting its potential to advance compact, fabrication-ready devices in integrated THz systems.

\section{Methods}

Traditional forward design methodologies for THz filters are largely rooted in analytical theories derived from microwave engineering. Typically, the design of periodic band-stop THz filters in planar waveguide structures begins by specifying a center frequency and target bandwidth, followed by estimating the filter period based on the effective relative permittivity of supported modes, usually obtained through full-wave simulations. The geometric parameters are then derived from this model and iteratively fine-tuned through additional simulations. Although this approach can yield functional designs, it inherently depends on simplifying assumptions and limited parameter sweeps, which constrain the design space and limit opportunities for performance optimization.

In contrast, inverse design frameworks allow users to define a desired spectral response, including both magnitude and phase characteristics, and automatically search for optimal geometries that meet those specifications. This enables the discovery of non-intuitive, high-performance structures that would be difficult to realize through conventional methods. Moreover, by formulating the design task as an optimization problem, inverse design reduces the reliance on trial-and-error tuning and provides a direct route to meeting stringent spectral requirements, particularly in compact and fabrication-constrained THz platforms.

In this work, we focus on the inverse design of optimized planar THz filters using a GA-based framework. Several synthesized band-stop filters are presented as proof-of-concept demonstrations to validate the proposed optimization approach. These filters feature varying center frequencies, bandwidths, and rejection depths, showcasing the flexibility and capability of the inverse design method. Band-stop filters were chosen as the initial test case to allow direct comparison with previous forward-designed filters, providing a meaningful benchmark for evaluating the performance improvements achieved through the GA-based inverse design process.

\subsection{Inverse design with genetic algorithm}

The GA offers several advantages that make it particularly suitable for the inverse design of complex electromagnetic structures. Unlike gradient-based optimization methods, GA does not require computation of gradients of the fitness (objective) function. This characteristic makes it more robust against local minima, which are common in non-convex high-dimensional design spaces \cite{haupt2007genetic}.

In this work, we employ a GA to optimize the binary distribution of gold (Au) and air pixels within a planar THz filter structure. Each candidate solution is encoded as a chromosome representing the spatial arrangement of pixels that evolve over successive generations through selection, crossover, and mutation. The GA parameters are carefully tuned to address the discrete, high-dimensional nature of the problem.

The mutation operator introduces diversity by toggling individual bits, corresponding to changing the material assignment of specific pixels. This allows for the exploration of novel topologies that can enhance the S-parameter response. Crossover combines favorable traits from two parent designs, while elitism ensures that top-performing individuals are preserved. The population size balances exploration breadth with computational feasibility, considering the cost of simulations~\cite{haupt2007genetic}.

Several strategies are employed to mitigate the risk of convergence to local minima that are common in nonconvex optimization landscapes defined by electromagnetic performance metrics. Increasing the mutation rate enhances diversity and allows broader sampling of the solution space. When a locally optimal design is suspected, its genome can be reintroduced in subsequent runs to allow refined exploration around its neighborhood. Extending the number of generations provides additional opportunities for the evolution of high-performance filter geometries. Additionally, running the GA multiple times with randomized initial populations allows for statistical comparison across runs, providing information on the robustness of the solutions and the sensitivity of the design problem to initial conditions.

\begin{figure*}[t]
    \centering  
    \includegraphics[trim=20 70 60 0, clip, width=\textwidth]{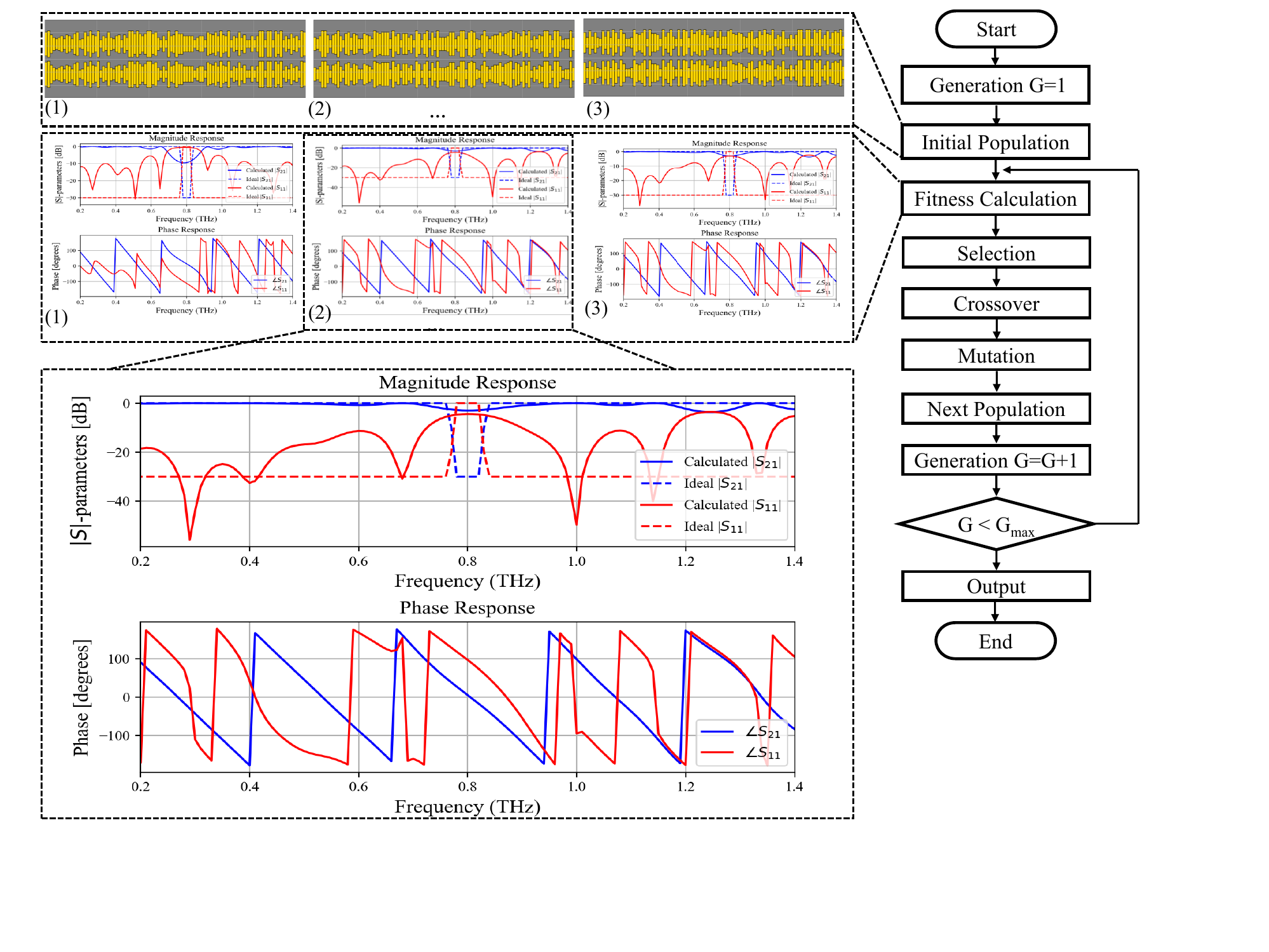}
    \caption{Flowchart of the GA optimization process. The randomly generated initial population of THz filters is depicted, along with their corresponding \( S \)-parameters (both magnitude and phase). The algorithm iterates through successive generations, refining the population until one of two stopping criteria is met: (1) the number of generations \( G \) reaches the predefined maximum \( G_{\max} \), or (2) the error function converges below a specified threshold.}
    \label{fig:TAMBG_result}
\end{figure*}

Figure~\ref{fig:TAMBG_result} presents the flow diagram of the GA optimization process that we used to design THz filters. The process begins with a randomly generated initial population of filters, each associated with its corresponding S-parameters, including both magnitude and phase. As the algorithm progresses, iteratively refines the population across successive generations based on performance feedback (fitness function). This evolutionary process enables the development of optimized THz filters that satisfy the specified design criteria.

We enforce structural continuity for compatibility with our THz test platform using a connectivity constraint that is incorporated into the design formulation. Specifically, the filter must maintain a continuous conductive pathway to facilitate voltage biasing of the transmitter photoconductive switch (PCS). Without such a constraint, generated structures may introduce discontinuities (i.e., a DC block), and the device would fail to operate correctly during experimentation. While this work does not include experimental validation, the design framework is developed with all necessary constraints in place to facilitate future experimental implementation.

With the structural constraints integrated into the initialization process, the inverse design problem is formally defined and formulated for optimization. Each candidate solution is represented as a binary matrix encoding the spatial distribution of gold and air pixels within the designated design region. The pixel size was set to 4~\textmu m$\times$10~\textmu m to balance resolution requirements, fabrication feasibility, and computational cost. This binary representation serves as input to the GA, which iteratively evolves the population to maximize the fitness function, defined as the figure of merit (FOM), that quantifies how closely the simulated electromagnetic response matches the S-parameters of the target filter.

\begin{figure*}[t]
  \centering
  \includegraphics[width=6in]{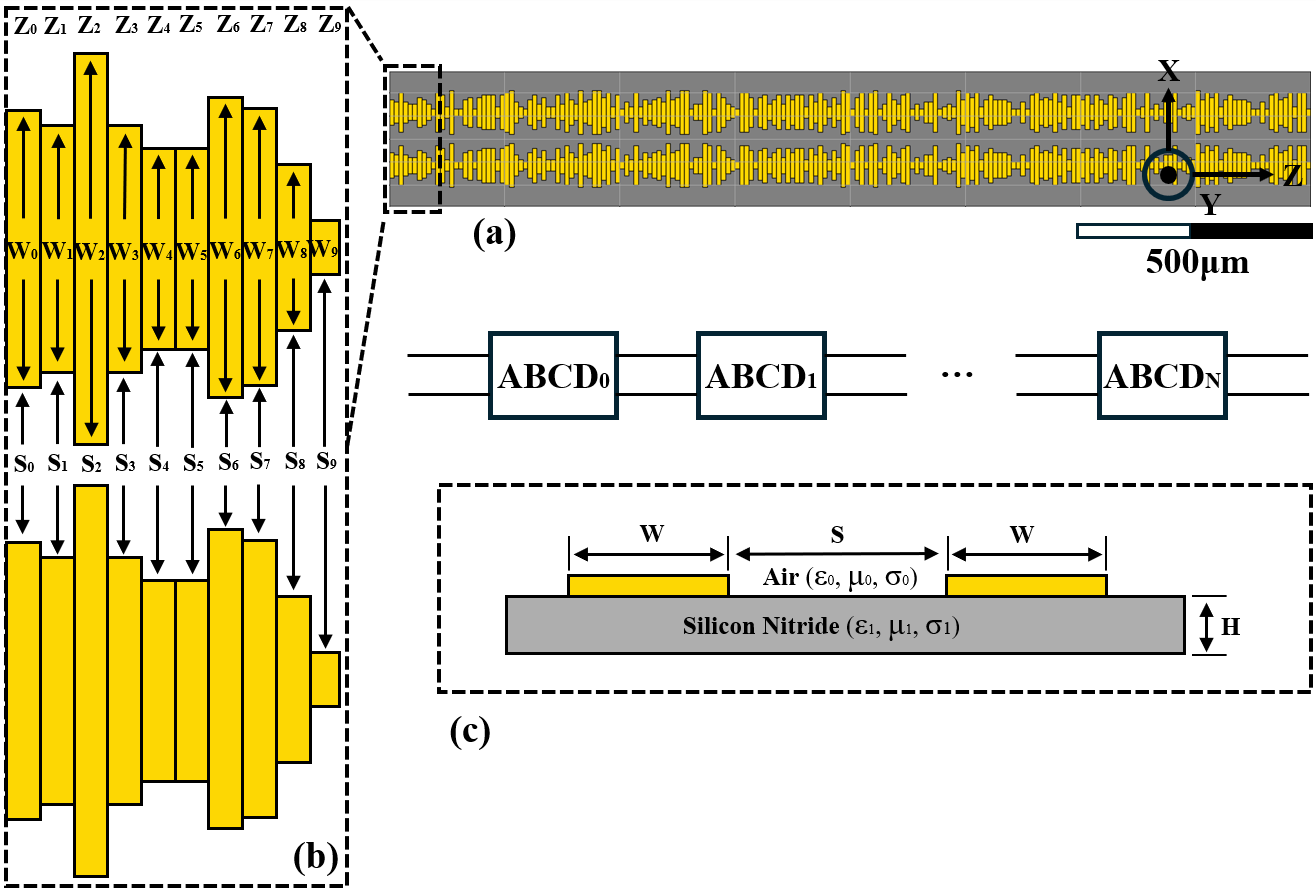}
\caption{
Illustration of the design framework and associated parameters. 
(a) Example of a planar THz filter structure synthesized using a GA-based inverse design approach. The overall framework dimensions are $300~\mu\mathrm{m} \times 2000~\mu\mathrm{m}$, corresponding to 200 discrete columns in the propagation direction. 
(b) Magnified view of a section of the filter, highlighting key design parameters: impedance ($Z$), conductor width ($W$), and strip spacing ($S$). 
(c) Cross-sectional schematic of the filter structure, depicting gold metallic layers patterned on a silicon nitride (Si$_3$N$_4$) membrane substrate.
}

  \label{fig:str}
\end{figure*}

Figure~\ref{fig:str} illustrates the design framework and key structural parameters of the planar THz filter investigated in this work. The filter region spans an area of 300~\textmu m in width and 2000~\textmu m in length, corresponding to 200 discrete columns along the propagation direction. This discretized domain serves as the optimization space for the GA-based inverse design process, wherein the binary configuration of each column defines the local characteristic impedance profile of the filter.

A magnified view of a representative section is shown in Fig.~\ref{fig:str}(b), highlighting essential design parameters, including the local characteristic impedance ($Z$), conductor width ($W$), and strip spacing ($S$). These geometrical features collectively govern the electromagnetic response of the structure, enabling fine-tuning of the filter’s spectral characteristics. In the design space explored, $W$ varies between 10~\textmu m and 90~\textmu m, while $S$ ranges from 5~\textmu m to 90~\textmu m, ensuring compatibility with fabrication constraints and impedance requirements.

Figure~\ref{fig:str}(c) presents a cross-sectional schematic of the filter, depicting gold metallic layers patterned on a silicon nitride (Si$_3$N$_4$) membrane substrate. This material platform ensures compatibility with standard microfabrication processes while minimizing substrate losses at THz frequencies. The combination of geometrical flexibility and material selection facilitates the realization of high-performance THz filters within a compact planar footprint.

After initializing the population, each candidate design is evaluated using the fitness function. The fitness function quantifies performance by calculating the root-mean-square error (RMSE) between the simulated and target scattering parameters S-parameters), incorporating both magnitude and phase.

Formally, the fitness function \( f(\mathbf{x}_i^t): \mathbb{R}^n \to \mathbb{R} \) maps design parameters \( \mathbf{x}_i^t \) at iteration \( t \) to a scalar score reflecting spectral deviation. The loss function \( \mathcal{L} = -\text{RMSE} \) guiding optimization is expressed as:

\begin{equation}
\begin{aligned}
\mathcal{L} = &\; w_1 \sqrt{\frac{1}{N} \sum_{i=1}^{N} (S_{21}^{\text{dB}}(f_i) - S_{21,\text{target}}^{\text{dB}}(f_i))^2} \\
& + w_2 \sqrt{\frac{1}{N} \sum_{i=1}^{N} (S_{11}^{\text{dB}}(f_i) - S_{11,\text{target}}^{\text{dB}}(f_i))^2} \\
& + w_3 \sqrt{\frac{1}{N} \sum_{i=1}^{N} (\angle S_{11}(f_i) - \angle S_{11,\text{target}}(f_i))^2} \\
& + w_4 \sqrt{\frac{1}{N} \sum_{i=1}^{N} (\angle S_{21}(f_i) - \angle S_{21,\text{target}}(f_i))^2}
\end{aligned}
\end{equation}

\noindent where \( N \) is the number of frequency points, and \( w_1, w_2, w_3, w_4 \) are weighting factors for magnitude and phase components of transmission (\( S_{21} \)) and reflection (\( S_{11} \)) responses. This formulation enables precise, frequency-aware tuning of THz filter designs through genetic optimization.

In this work, the target phase response was specified as linear; therefore, \( w_3 \) and \( w_4 \) were assigned to 10\% of the total loss function weight. As a result, the optimization process initially prioritizes minimizing the magnitude-related components (\( w_1 \) and \( w_2 \)) of the loss function, with phase error contributing less strongly during early optimization. In the later stages, the algorithm fine-tunes the design by evaluating the differential phase between adjacent frequency points and reducing phase variations through the influence of \( w_3 \) and \( w_4 \).

A population size of 200 was selected to balance exploration breadth with computational feasibility. Preliminary experiments conducted with population sizes of 50, 100, 200, and 400 showed diminishing fitness improvements beyond 200, while computational time scaled proportionally. Each generation evolves through a combination of elitism, selection, crossover, and mutation, with the objective of maximizing the fitness function.

To guide the evolutionary process toward optimal solutions, an elite group comprising the top 30 individuals (15\% of the population) with the highest fitness values is directly carried over to the next generation. This elite fraction was empirically chosen to balance convergence stability and exploration diversity: higher elite rates ($>20\%$) accelerated convergence but increased the risk of premature convergence to local optima, while lower rates ($<10\%$) slowed convergence without yielding notable diversity gains. The elite individuals are identified by simulating all 200 candidates, ranking them by fitness, and preserving the top 30.

The remaining 170 candidates are generated through the application of genetic operators: selection, crossover, and mutation. To maintain population diversity and promote exploration of the design space, a probabilistic rank-based tournament selection strategy is employed. In this method, a subset of \( T \) individuals is randomly sampled from the population, and an individual is selected from the tournament with a probability that favors higher-ranked individuals. The probability of selecting an individual \( i \) with rank \( r_i \) (where \( r = 1 \) corresponds to the highest fitness) is given by:

\begin{equation}
P(i) = \frac{(1 - s) s^{r_i - 1}}{1 - s^T}
\end{equation}

\noindent where \( s \in (0,1) \) is the selection pressure and \( T \) is the tournament size~\cite{miller1995genetic, rawlins2014foundations}. This probabilistic approach allows lower-ranked individuals to be selected with nonzero probability, promoting genetic diversity while still biasing selection toward fitter individuals.

A tournament size of \( T = 4 \) was selected based on literature recommendations for moderate selection pressure in combinatorial GAs~\cite{haupt2007genetic, miller1995genetic}. Preliminary tests showed that smaller \( T \) slowed convergence, while larger \( T \) reduced diversity. This value was empirically validated to achieve consistent fitness improvements across runs, balancing exploitation and exploration and reducing the risk of premature convergence.

\begin{figure*}[t]
    \centering  
    \includegraphics[width=\textwidth]{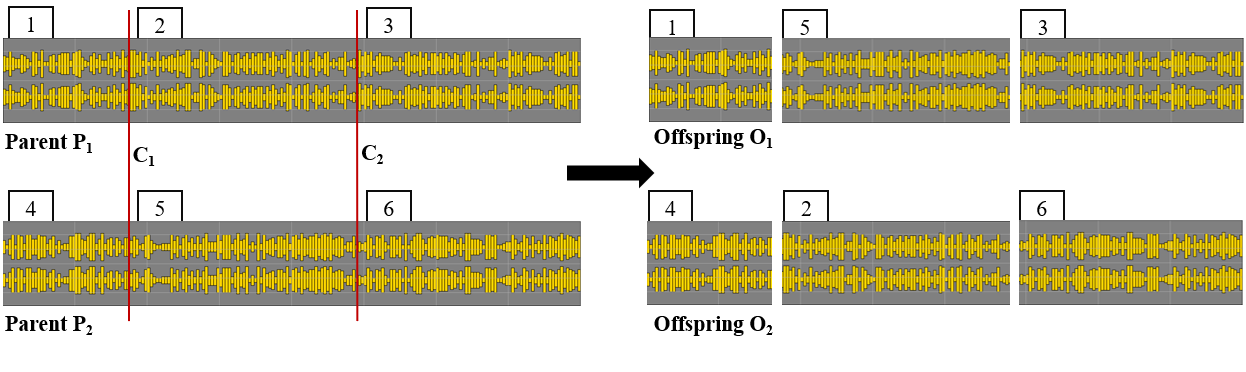}
    \caption{Illustration of the two-point crossover mechanism employed in the GA-based inverse design of THz filters. This crossover process facilitates the recombination of structural features from two-parent designs, generating offspring with inherited characteristics that contribute to the optimization of the filter’s electromagnetic response.}
    \label{fig:Crossover}
\end{figure*}

Crossover is employed as a recombination operator to generate new candidate solutions by combining structural characteristics from two parent individuals. In the context of grid-based THz filter design, a two-point crossover strategy is implemented to facilitate the creation of novel geometries while preserving connectivity and material constraints (Fig.~\ref{fig:Crossover}). Given two parent grids \( P_1 \) and \( P_2 \), two crossover points \( c_1 \) and \( c_2 \) are randomly selected at uniformly distributed positions within the grid such that \( 2 \leq c_1 < c_2 \leq C - 1 \), where \( C \) represents the total number of columns in the grid. Two offspring, \( O_1 \) and \( O_2 \), are then generated by exchanging the column segments between \( c_1 \) and \( c_2 \):

\begin{equation}
O_1 = [P_1(:, 1:c_1),\; P_2(:, c_1:c_2),\; P_1(:, c_2:C)]
\end{equation}

\begin{equation}
O_2 = [P_2(:, 1:c_1),\; P_1(:, c_1:c_2),\; P_2(:, c_2:C)]
\end{equation}

\noindent Here, \( P(:, a:b) \) denotes the submatrix containing columns \( a \) to \( b \). The use of two crossover points allows for greater structural variation compared to single-point schemes, while maintaining design continuity and compatibility with the CPS platform.

To further enhance diversity, mutation is applied to 10\% of the selected individuals. During mutation, a randomly chosen bit in the binary chromosome is flipped (i.e., ``0'' to ``1'' or ``1'' to ``0''), corresponding to a material change at a specific pixel—effectively toggling it between a gold and air region. This stochastic perturbation introduces novel traits into the population and supports broader exploration of the design space. Importantly, all mutation operations are performed while enforcing the continuity constraint, ensuring that conductive pathways remain uninterrupted to maintain compatibility with the CPS platform in future experiments.

\section{SIMULATION RESULTS AND DISCUSSION}

To compute the S-parameters and subsequently evaluate the RMSE, two computational methodologies are available: (1) an analytical approach based on the ABCD transmission matrix method, and (2) a full-wave numerical analysis performed using commercial software such as ANSYS HFSS. Given the substantial computational cost associated with FEM-based simulations, the ABCD transmission matrix method was employed as an efficient surrogate modeling approach. This method provides a computational speed-up exceeding three orders of magnitude compared to full-wave simulations, thereby enabling rapid evaluation of design candidates during the optimization process while maintaining sufficient accuracy for guiding the inverse design. Prior studies have demonstrated a strong correlation between the results obtained from this analytical approach and experimental measurements, thereby validating its accuracy~\cite{frankel1991terahertz}.

The designed filters are composed of cascaded unit cells, each modeled as a section of transmission line with a specific characteristic impedance. These unit cells are arranged laterally along the column axis of the design grid, with each column corresponding to one such unit cell. To accurately model the behavior of these structures, each unit cell is represented by a two-port ABCD matrix, which characterizes the voltage and current relationship at its input and output ports. This modeling approach assumes that each section behaves as a uniform transmission line, which is a valid approximation under the quasi-static (QS) regime—where the physical dimensions of the unit cells are much smaller than the operating wavelength.

The ABCD matrix of a transmission line segment of length \( \Lambda \), characteristic impedance \( Z \), and phase constant \( \beta \) is given by:

\begin{equation}
\begin{bmatrix}
A_{\text{cell}} & B_{\text{cell}} \\
C_{\text{cell}} & D_{\text{cell}} 
\end{bmatrix} =
\begin{bmatrix}
\cos(\beta \Lambda) & jZ \sin(\beta \Lambda) \\
j \frac{1}{Z} \sin(\beta \Lambda) & \cos(\beta \Lambda)
\end{bmatrix}
\label{eqn:ABCD_cell}
\end{equation}

The overall ABCD matrix for a filter composed of \( N \) cascaded unit cells is obtained by sequentially multiplying the individual ABCD matrices corresponding to each unit cell:

\begin{equation}
\begin{bmatrix}
A & B \\
C & D 
\end{bmatrix}_{\text{total}} =
\begin{bmatrix}
A_{C1} & B_{C1} \\
C_{C1} & D_{C1} 
\end{bmatrix}
\cdot
\begin{bmatrix}
A_{C2} & B_{C2} \\
C_{C2} & D_{C2} 
\end{bmatrix}
\cdots
\begin{bmatrix}
A_{CN} & B_{CN} \\
C_{CN} & D_{CN} 
\end{bmatrix}
\label{eqn:ABCD_total}
\end{equation}

Each section's QS characteristic impedance is computed analytically based on its lateral geometry, including the conductor width \( W \) and the inner strip spacing \( S \), using the closed-form expression for CPS transmission lines \cite{ghione1984analytical}:

\begin{equation}
Z_{\text{CPS}} = \frac{120\pi}{\sqrt{\varepsilon_{re}}} \frac{K(k_{\text{CPS}})}{K(k_{\text{CPS}}')}, \quad k_{\text{CPS}} = \frac{S}{S+2W}
\label{eqn:ZCPS}
\end{equation}

\noindent where \( K(k) \) and \( K(k') \) are the complete elliptic integrals of the first kind, and \( \varepsilon_{re} \) is the effective relative permittivity of the substrate.

Finally, the scattering parameters (S-parameters) are extracted from the total ABCD matrix using standard transformations~\cite{pozar2021microwave}:

\begin{equation}
S_{11} = \frac{A + \frac{B}{Z_{\text{CPS}}} - CZ_{\text{CPS}} - D}{A + \frac{B}{Z_{\text{CPS}}} + CZ_{\text{CPS}} + D}, 
S_{21} = \frac{2}{A + \frac{B}{Z_{\text{CPS}}} + CZ_{\text{CPS}} + D}
\label{eqn:ABCD_to_S}
\end{equation}

This modeling framework enables rapid and reasonably accurate prediction of the electromagnetic response of cascaded THz filters, making it well-suited for evaluating the fitness of a large number of candidate structures during the inverse design process, particularly in early optimization stages where full-wave simulations would be computationally prohibitive.

\begin{figure*}[t]
    \centering  
    \includegraphics[trim=0 120 20 0, clip, width=6in]{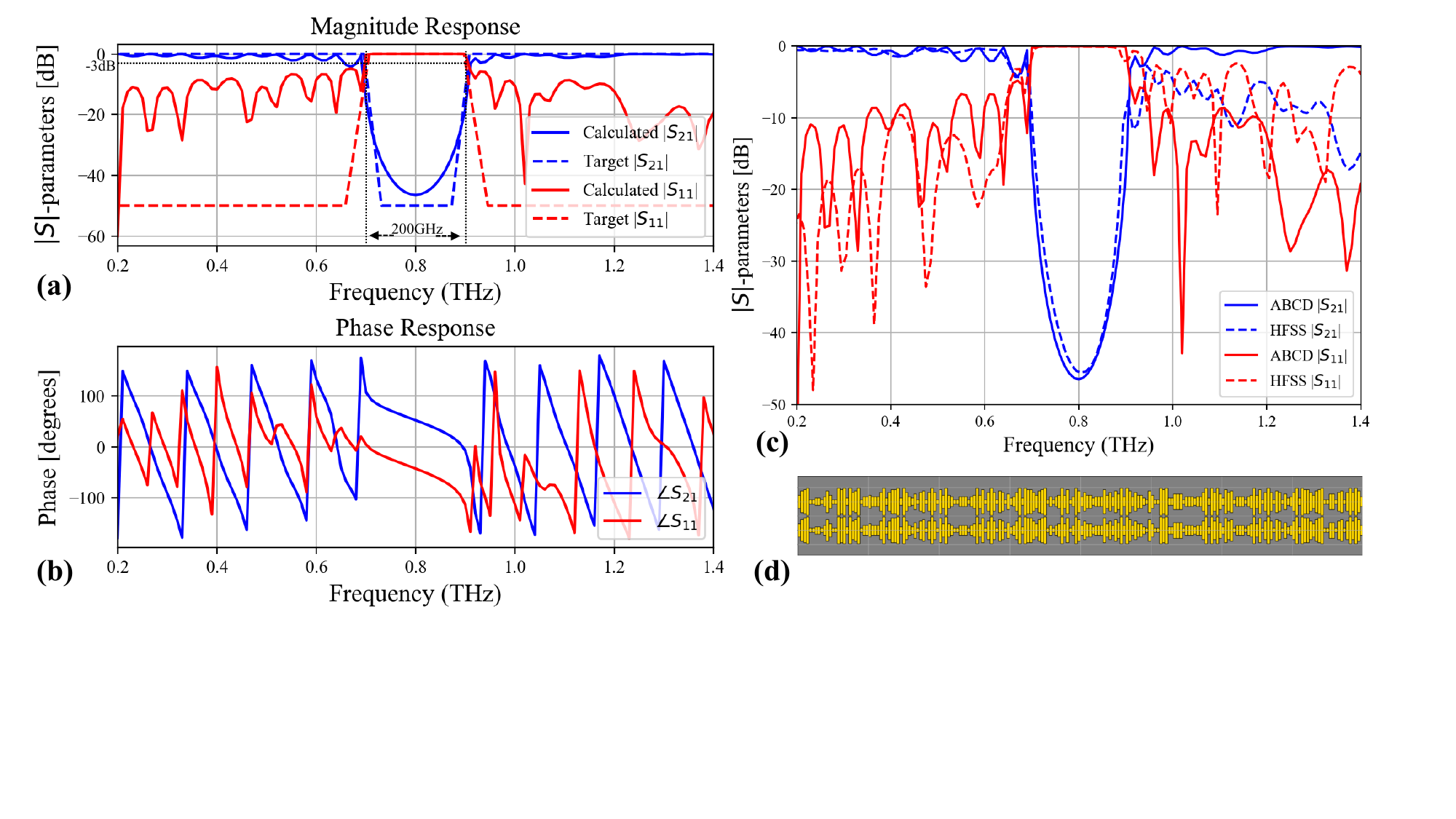}
    \caption{
    (a) Calculated magnitude response of the planar band-stop filter's S-parameters ($S_{21}$ and $S_{11}$) using the ABCD matrix method, compared against the target response. The design specifies a 3-dB bandwidth of 200~GHz and a rejection level of $-50$~dB, with the computed response achieving a rejection depth of approximately $-46$~dB.
    (b) Phase response of the calculated S-parameters, emphasizing the linear phase progression of $S_{21}$ across the passband.
    (c) Comparison between the results predicted by the ABCD matrix method and full-wave electromagnetic simulations using ANSYS HFSS, showing strong agreement.
    (d) Final optimized filter geometry synthesized using the proposed GA-based inverse design framework.
    }
    \label{fig:s-param-final}
\end{figure*}

Figure~\ref{fig:s-param-final} summarizes the spectral response and structural realization of a planar band-stop filter synthesized using the proposed GA-based inverse design framework. The target specifications include a center frequency of 0.8~THz and a 3-dB bandwidth of 200~GHz. In Fig.~\ref{fig:s-param-final}(a), the calculated magnitude response of $S_{21}$ and $S_{11}$, obtained via the ABCD matrix method, is compared against the target response. The design meets the specified rejection depth and bandwidth with high accuracy. Figure~\ref{fig:s-param-final}(b) presents the corresponding phase response, with particular emphasis on the linear phase progression of $S_{21}$ across the passband, which is critical for dispersion-sensitive THz applications. A direct comparison between the ABCD matrix-based results and full-wave electromagnetic simulations from ANSYS HFSS is provided in Fig.~\ref{fig:s-param-final}(c). Strong agreement is observed across most of the frequency range, validating the modeling framework. However, minor discrepancies emerge at higher frequencies (above 0.9~THz), where the ABCD matrix method begins to diverge from HFSS simulation. These deviations are primarily attributed to radiation losses and edge diffraction effects that are not captured by the quasi-static assumptions inherent in the ABCD matrix model. Lastly, Fig.~\ref{fig:s-param-final}(d) depicts the unconventional filter geometry synthesized through the GA optimization process, illustrating the method’s ability to generate non-intuitive yet high-performance designs within the imposed fabrication constraints.

\begin{figure*}[t]
    \centering  
    \includegraphics[trim=100 80 60 20, clip, width=5.3in]{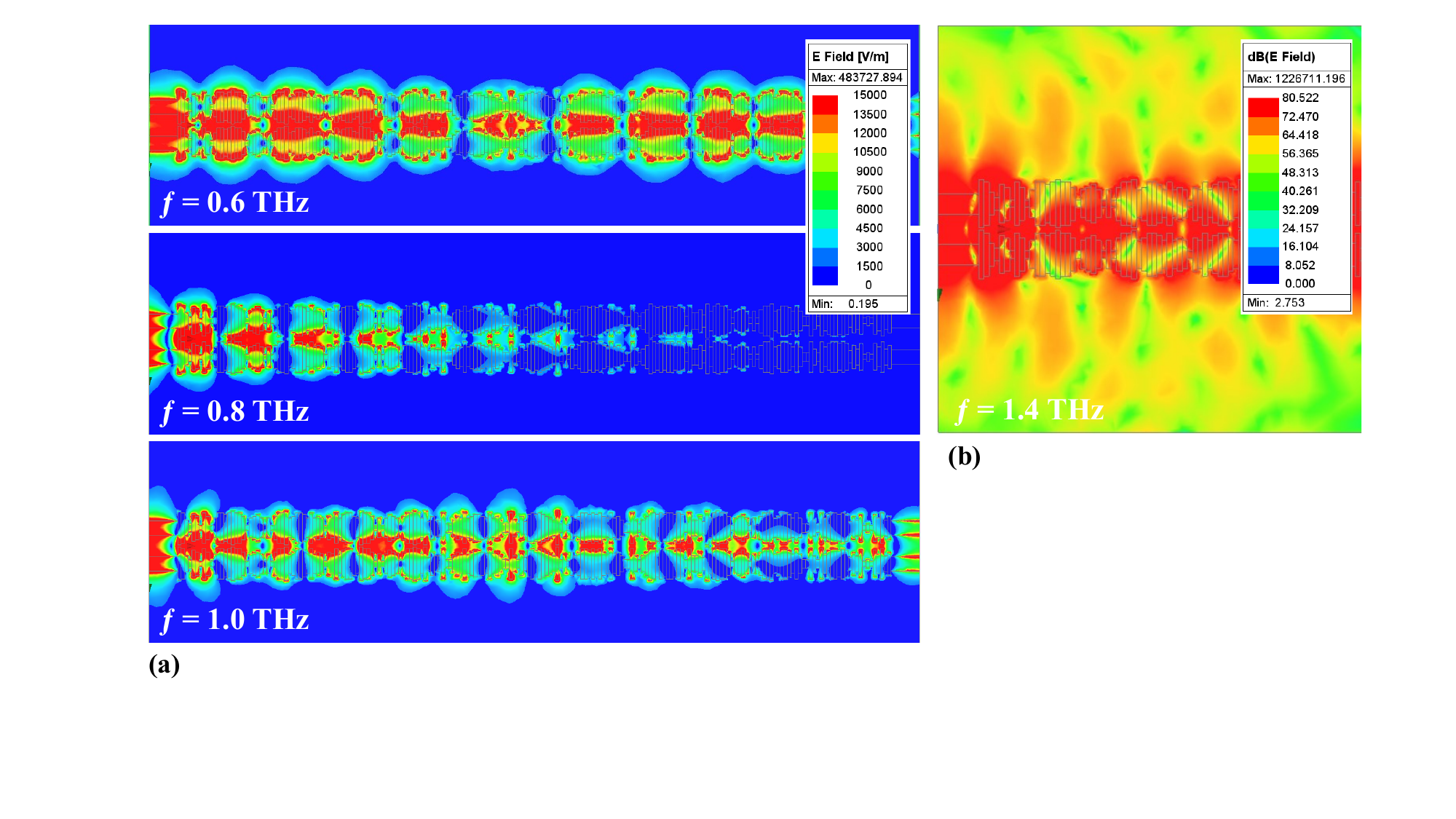}
    \caption{
    (a) Electric field distribution of the inverse-designed planar band-stop filter at 0.6~THz, 0.8~THz, and 1.0~THz. The filter exhibits strong suppression in the stopband (0.8~THz) and effective transmission in the passbands (0.6~THz and 1.0~THz), validating its spectral selectivity. 
    (b) Logarithmic-scale field profile at 1.4~THz, illustrating radiation losses at higher frequencies due to edge diffraction and imperfect confinement, particularly from the conductor edges.
    }
    \label{fig:E_field}
\end{figure*}

To further investigate the electromagnetic behavior of the optimized filter structure, the electric field distribution was analyzed at multiple frequencies spanning the passband and stopband. Figure~\ref{fig:E_field}(a) shows the simulated in-plane $E$-field profiles at 0.6~THz, 0.8~THz, and 1.0~THz. At 0.6~THz and 1.0~THz, the field propagates effectively through the structure, confirming low insertion loss in the passbands. In contrast, at 0.8~THz—corresponding to the filter's stopband—strong field attenuation is observed, demonstrating effective suppression of transmission and validating the spectral selectivity of the inverse-designed filter.

Figure~\ref{fig:E_field}(b) presents the field profile at 1.4~THz on a logarithmic scale to highlight radiation effects at higher frequencies. The simulation reveals radiation leakage originating from the edges of the metallic conductors, attributed to diffraction at frequencies beyond the designed stopband.

\begin{figure}[h]
    \includegraphics[trim=150 100 200 20, clip, width=3.5in]{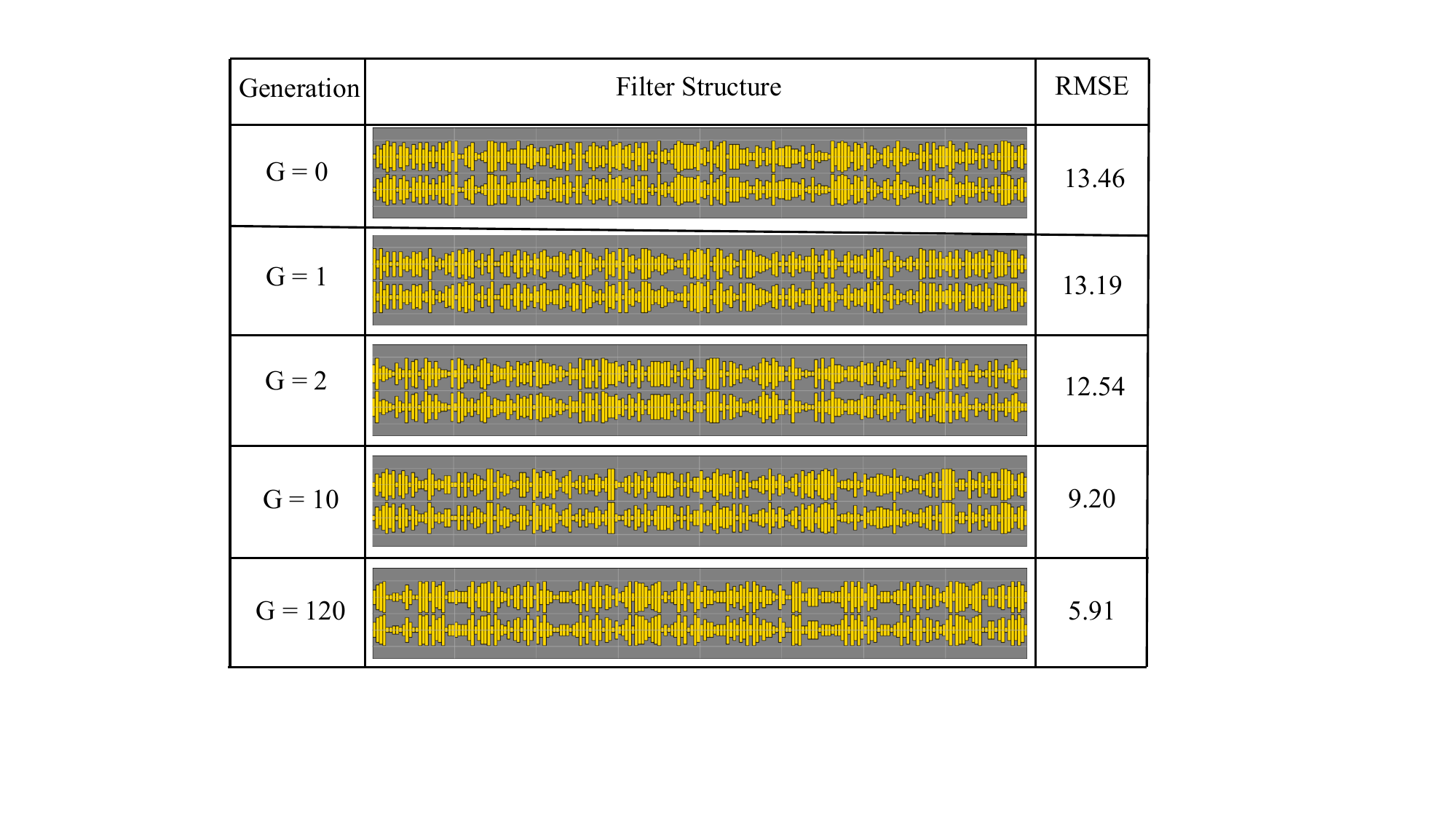}
    \caption{
    (e) Evolutionary optimization process for the inverse-designed planar THz band-stop filter. The initial generation begins with a randomly selected structure, which is iteratively refined across successive generations. The algorithm progressively minimizes the error function until, after 120 generations, the design converges to the desired target filter configuration, demonstrating the efficacy of the proposed inverse design framework.
    }
    \label{fig:GA_evolution}
\end{figure}

Figure~\ref{fig:GA_evolution} illustrates the evolutionary optimization process over multiple generations. The initial generation consists of randomly generated structures, which are progressively refined as the algorithm minimizes the error function. After 120 generations, the optimization converges to a design that matches the desired band stop filter response, as evidenced by the corresponding S-parameters (\( S_{21} \) and \( S_{11} \) in both magnitude and phase).

\begin{figure*}[t]
    \centering  
    \includegraphics[trim=0 20 10 60, clip, width=7.2in]{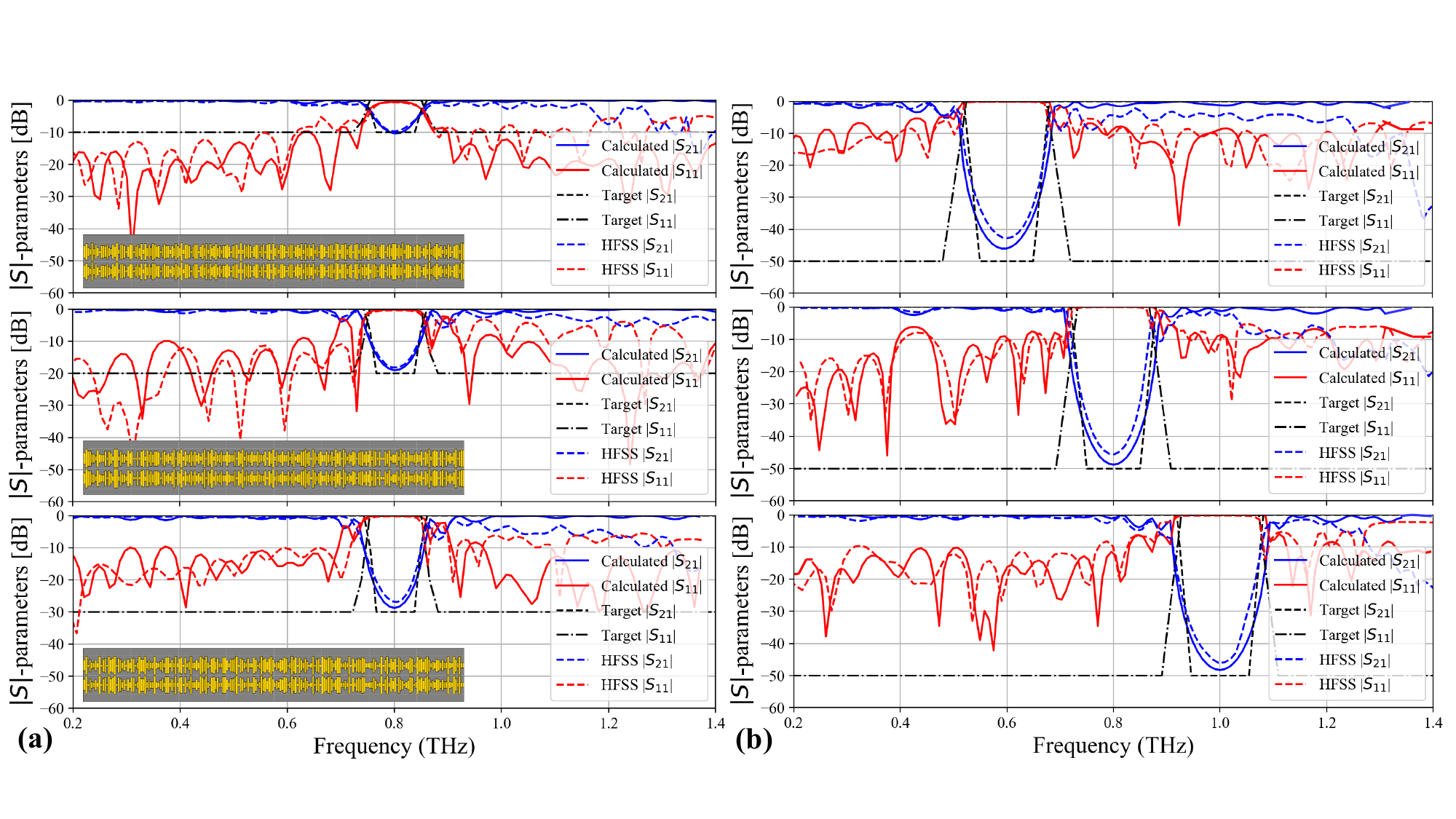}
    \caption{Comparison of S-parameters of band-stop filters designed using the GA-based inverse design framework. 
    (a) Filters with varying rejection depths centered at 0.8~THz, all within the same device footprint. 
    (b) Filters targeting different center frequencies of 0.6~THz, 0.8~THz, and 1.0~THz, respectively, synthesized within the same device footprint. 
    In both cases, S-parameters obtained using the ABCD matrix method are compared with full-wave simulations performed in ANSYS HFSS, demonstrating strong agreement.}
    \label{fig:results}
\end{figure*}

To demonstrate the capabilities of the proposed inverse design framework, two sets of design objectives were considered to systematically evaluate the ability of the framework to achieve distinct spectral responses while maintaining a fixed physical footprint of 2000~\textmu m in length. In the first set, we investigated the system's ability to control the rejection depth without altering the device geometry. Specifically, the GA-based inverse design framework was tasked to synthesize band-stop filters centered at 0.8~THz with a fixed 3-dB bandwidth of 70~GHz, while achieving three distinct rejection depths of 10~dB, 20~dB, and 30~dB, respectively. This evaluation serves as a critical benchmark of the algorithm’s capability to precisely tailor the depth of the stopband response by optimizing the internal geometry, despite the fixed device dimensions and fabrication constraints. Figure~\ref{fig:results}(a) presents the simulated S-parameters of the designed filters, demonstrating the system's flexibility in producing filters with different rejection depths. The results confirm that, through appropriate geometry optimization guided by the RMSE-based fitness function, the framework successfully generates non-intuitive filter structures capable of meeting the specified rejection levels. Compared to conventional Bragg grating filters of similar spectral specifications~\cite{dehghanian2025demonstration}, the GA-based inverse-designed filters achieve superior performance (-30~dB rejection compare to -18~dB) with approximately 50\% reduction in device length, underscoring the effectiveness of the proposed method for compact THz filter synthesis.

In the second set, we explored the framework’s capability to achieve center frequency tunability within the same physical platform and device length. Band-stop filters with center frequencies of 0.6~THz, 0.8~THz, and 1.0~THz, each targeting a 3-dB bandwidth of 150~GHz, were synthesized. This evaluation demonstrates the adaptability of the inverse design framework to achieve wide frequency tunability by modifying the target spectral response in the optimization process, without requiring changes to the overall device footprint. The simulated S-parameters of these filters are shown in Fig.~\ref{fig:results}(b). These findings highlight the ability of the proposed framework to design frequency-selective filters with distinct geometries, adapted to various operational frequencies, while preserving a consistent physical footprint and ensuring compatibility with existing fabrication constraints. This underlines the versatility of the developed system for creating customized THz filters capable of meeting diverse application requirements in integrated platforms.

The comparison between the results obtained from the ABCD matrix method and full-wave electromagnetic simulations is presented in each figure. The two approaches exhibit strong agreement with minimal divergence across and below the operational bandwidth. Minor discrepancies at higher frequencies can be attributed to radiation losses and edge diffraction effects, which are not captured by the quasi-static assumptions of the ABCD matrix model, as discussed previously.

The integration of the ABCD matrix method into the GA-based inverse design framework provided a critical computational advantage, enabling rapid and efficient evaluation of candidate filter structures. Each optimization run—consisting of 120 generations and a population size of 200—was completed in approximately 40 minutes on an Intel i7-10700 CPU. In contrast, a single full-wave simulation in ANSYS HFSS (with a maximum $\Delta S = 0.02$ and 130 frequency points) required over two hours, making large-scale optimization infeasible without a surrogate model. The use of the ABCD method thus enabled extensive design space exploration and facilitated the discovery of non-intuitive, high-performance filter geometries compatible with planar CPS-integrated platforms.

\begin{figure*}[t]
    \centering
    \includegraphics[trim=0 220 60 0, clip, width=6in]{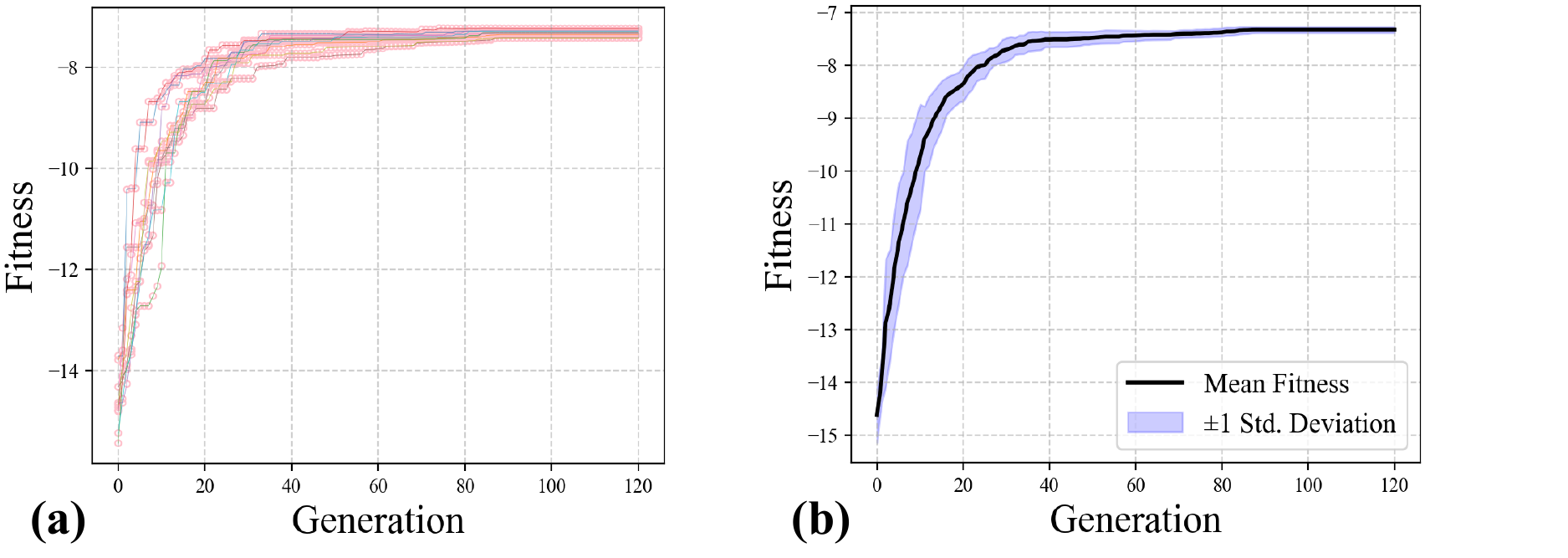}
    \caption{
    Convergence behavior of the GA-based inverse design framework over 10 independent runs initialized with different random seeds. 
    (a) Each ring marker represents the best fitness value achieved in that generation for a single run. 
    (b) The solid blue curve indicates the average best fitness across generations, while the shaded region denotes the $\pm1$ standard deviation range. 
    The results demonstrate consistent convergence behavior, low variance across runs, and robustness with respect to initial population randomness.
    }
    \label{fig:convergence_10runs}
\end{figure*}

Convergence analysis was performed to assess the stability and consistency of the GA-based inverse design framework. As illustrated in Fig.~\ref{fig:convergence_10runs}, the algorithm was independently executed ten times, each with a different random seed to evaluate the sensitivity of the optimization process to initial population variation. Figure~\ref{fig:convergence_10runs}(a) visualizes the best fitness value at each generation for all ten runs using circular ring markers. The overlapping distribution of fitness values across runs highlights the consistent convergence trajectory and low inter-run variability.

Figure~\ref{fig:convergence_10runs}(b) presents a statistical summary of the convergence behavior. The solid blue curve represents the average best fitness value across the ten runs at each generation, while the shaded region indicates the $\pm1$ standard deviation range. The convergence curve exhibits a rapid increase in fitness during the first 40–50 generations, followed by a slower refinement phase, with convergence typically reached between generations 100 and 120. The narrow standard deviation—remaining below 3-dB in the final $S_{21}$ magnitude response—demonstrates high repeatability and convergence stability.

Although a no-improvement stopping condition was implemented, defined as halting the run if no progress was observed for 30 consecutive generations, it was not triggered in any of the cases. This behavior confirms that the optimization process continued to explore the design space effectively until the predefined maximum generation limit, ensuring thorough solution refinement without early stagnation.

Beyond conventional parameter sweep-based methods, the proposed approach enables automated, efficient, and flexible synthesis of THz filters by iteratively refining candidate solutions in high-dimensional, nonconvex design spaces. The adaptability of the framework allows it to accommodate diverse design objectives, including varying center frequencies, rejection depths, and impedance matching requirements, making it a promising tool for next-generation TSoC components. Future extensions will expand the framework's applicability to other planar devices such as couplers, reflectors, and absorbers.

\section{Conclusion}

We presented a GA-based inverse design framework for synthesizing planar THz band-stop filters integrated with coplanar stripline (CPS) transmission lines. The framework enabled efficient exploration of complex design spaces, generating non-intuitive, high-performance filter geometries. To accelerate the optimization process, the ABCD matrix method was employed, offering a substantial speed-up over full-wave simulations, with final validation performed using FEM-based electromagnetic analysis. The framework successfully demonstrated the design of filters with tunable rejection depths and center frequencies within fixed device footprints. This work represents the first application of GA-based inverse design using the ABCD matrix method for CPS-integrated THz filters and highlights its potential to advance compact, fabrication-ready components in terahertz system-on-chip (TSoC) platforms.

\section*{Acknowledgments}
This work was supported by a NSERC Discovery Grant.

This article used Writeful in Overleaf for editing purposes.

\bibliographystyle{IEEEtran}
\bibliography{IEEEabrv,Reference}

\vfill

\end{document}